# Eddington, Lemaître and the hypothesis of cosmic expansion in 1927

Cormac O'Raifeartaigh

*School of Science and Computing, Waterford Institute of Technology, Cork Road, Waterford, Ireland*

Author for correspondence: coraifeartaigh@wit.ie

# 1. Introduction

Arthur Stanley Eddington was one of the leading astronomers and theorists of his generation (Smart et al. 1945; McCrea 1982; Chandrasekhar, S. 1983). An early and important proponent of the general theory of relativity, his 1918 '*Report on the Relativity Theory of Gravitation*' for the Physical Society (Eddington 1918) provided an early authoritative exposition of the subject for English-speaking physicists (Vibert Douglas 1956 p42; Chandrasekhar 1983 p24). He played a leading role in the eclipse observations of 1919 that offered early astronomical evidence in support of the theory (Vibert Douglas 1956 pp 39-41; Chandrasekhar 1983 pp 24-29; Kennefick, this volume) while his book '*Space, Time and Gravitation*' (Eddington 1920) was one of the first popular treatises on general relativity for an English-speaking audience. In addition, Eddington's textbook '*The Mathematical Theory of Relativity*' (Eddington 1923) became a classic reference for English-speaking physicists with an interest in relativity (McCrea 1982; Chandrasekhar 1983 p32). Indeed, the book provided one of the first textbook accounts of relativistic models of the cosmos, complete with a discussion of possible links to one of the greatest astronomical puzzles of the age, the redshifts of the spiral nebulae (Eddington 1923 pp 155-170).

It seems therefore quite surprising that, when Eddington's former student Georges Lemaître suggested in a seminal article of 1927 (Lemaître 1927) that a universe of expanding radius could be derived from general relativity, and that the phenomenon could provide a natural explanation for the redshifts of the spiral nebulae, Eddington (and others) paid no attention. The oversight is particularly puzzling as it is known that Eddington received a copy of the paper from its author at the time of publication, as detailed below. Three years later, Eddington embraced Lemaître's hypothesis with great enthusiasm, distributing it to colleagues and arranging for it to be republished in the *Monthly Notices of the Royal Astronomical Society*.

In this article, we consider the reasons Lemaître's hypothesis of cosmic expansion was overlooked by Eddington and others when it was first proposed. We find that conventional explanations for the oversight (such as Lemaître's status as a relatively junior researcher and his decision to publish in a lesser-known Belgian journal) do not convince. We consider an alternative explanation that has not been considered in the literature – namely that the astronomical data cited by Lemaître in his 1927 article in support of his model of cosmic expansion were not sufficiently robust to cause Eddington and others to take the model seriously.

## 2. Historical context: a paradigm shift delayed

On January 10th 1930, a landmark meeting took place at the Royal Astronomical Society in Burlington House in London. It was noted that recent observations by astronomers of an approximately linear relation between the redshifts of the spiral nebulae and their radial distances could not be readily explained in the context of the standard mathematical models of the cosmos, i.e., in the context of the static cosmology of Albert Einstein (Einstein 1917) or the empty cosmology of Willem de Sitter (de Sitter 1917). During the course of the meeting, it was suggested that non-static cosmologies should be considered (Kragh 1996 p21, 31; Nussbaumer and Bieri 2009 p121). This discussion was reported in the February issue of *The Observatory* (de Sitter 1930) and read by Georges Lemaître, who immediately wrote a letter to Eddington reminding him of his 1927 article on this very topic. As Lemaître wrote: *"Dear Professor Eddington, I just read the February $N^0$ of the Observatory and your suggestion of investigating of non-statical intermediary solutions between those of Einstein and de Sitter. I made these investigations two years ago. I consider a universe of curvature constant in space but increasing in time. And I emphasize the existence of a solution in which the motion of the nebulae is always a receding one from time minus infinity to plus infinity"* (Lemaître 1930). After giving some details of his cosmological model, Lemaître concluded the letter by remarking: *"I send you a few copies of the paper. Perhaps you find occasion to give it to de Sitter. I sent him also at the time but probably he didn't read it"*.[1]

As is well-known, Eddington responded with enthusiasm, publicly acknowledging the importance of Lemaître's 1927 article (Eddington 1930) and bringing it to the attention of his colleagues. Eddington also arranged for the article to be republished in English in the *Monthly Notices of the Royal Astronomical Society* (Kragh 1996 p32; Nussbaumer and Bieri 2009 pp 122-125).

Many years later, the British astronomer George McVittie, a former research student in Eddington's group, recalled the day that Eddington received Lemaître's letter: *"I remember the day when Eddington, rather shamefacedly, showed me a letter from Lemaître which reminded Eddington of the solution to the problem which Lemaître had already given. Eddington confessed that although he had seen Lemaître's paper in 1927 he had forgotten completely about it until that moment"* (McVittie 1967). A second account was provided by

---
[1] The letter can be read in full in (Nussbaumer and Bieri 2009) pp 122-123.

McVittie in oral testimony to the historian David DeVorkin in 1978. He recalled Eddington saying: *"I'm sure Lemaître must have sent me a reprint, he's just sent me another, but I'd forgotten about it"* (McVittie 1978; Kragh 1996 p32). Both statements neatly encapsulate our puzzle; given Eddington's great interest and expertise in both astronomy and relativity, it seems very strange that he had "forgotten" a paper offering an explanation for the redshifts of the nebulae in the context of a relativistic expansion of space.

3. **Conventional explanations**

   *(i)    Status of the researcher*

It is often noted that, in comparison with famous astronomers such as Eddington and Hubble, Lemaître was a relatively junior researcher when his seminal article of 1927 was published and it is certainly true that busy professors do not always take the time to read every new paper by a junior researcher. More generally, a daring new scientific hypothesis tends to be more readily accepted if it is proposed by an eminent scientist rather than a comparative unknown (Merton 1973 p443; Zuckerman 1973 pp 96-144). Thus, it is sometimes inferred that Lemaître's paper may not have been taken seriously due to his relatively junior status in the world of academia [van der Bergh 2011a; Grøn 2018].

However, this explanation overlooks the fact that Lemaître spent a year as a Research Associate at the University of Cambridge under the supervision of Eddington. It is known that the young cleric made a very favourable impression on his supervisor; indeed Eddington recommended Lemaître as *"a brilliant student … of great mathematical ability"* (Mitton 2017). We recall also that just a few years later, Lemaître achieved an important theoretical advance with a paper that demonstrated a significant flaw in de Sitter's cosmology (Lemaître 1925). This work received some attention and undoubtedly established the young cleric as a relativist of note (Mitton 2017; Kragh 2018).

 *(ii) Status of the journal*

It has also been suggested that the journal in which Lemaître's paper of 1927 was published, the *Annales de la Société Scientifique de Bruxelles*, was a rather obscure vehicle for publication (Nussbaumer and Bieri 2009 p99; van den Bergh, S. 2011a,b; Ostriker and Mitton 2013 p68; Kragh 2018). However, recent scholarship suggests that this point may have been somewhat overstated; in fact the *Annales* was a respected journal that could be found in the libraries of most European universities at the time. Articles in both French and English could be found in the journal and it was reasonably well-known throughout European academia (Lambert 2015 pp 132-133; Luminet 2013). In any case, it is known that Lemaître sent his paper directly to Eddington (and to de Sitter) and thus the status and circulation of the *Annales* is of little relevance.

It has similarly been suggested that the publication of Lemaître's article in French, rather than English, may have impacted the reception of the paper (van der Bergh 2011a; Ostriker and Mitton 2013 p68; Kragh 2018). However, it was hardly unusual for English-speaking scientists to read scientific papers published in French at this time. Moreover, Eddington himself translated a number of articles and books on relativity from French to English in these years (Smart et al. 1945; Laguens, this volume). We also note that the title of Lemaître's article was extremely specific. The first phrase, '*Un Univers Homogène de Masse Constante et de Rayon Croissant*' makes clear the principal topic of the paper (*A Homogeneous Universe of Constant Mass and Increasing Radius*) even to a reader not fluent in French. Similarly, the second phrase, '*Rendant Compte de la Vitesses Radiale des Nébeleuses Extra-Galactiques*' highlights the link to astronomy (*Accounting for the Radial Speeds of the Extra-Galactic Nebulae*). Thus, the entire contents of Lemaître's article of 1927 are summarized in the title, at least to readers with a basic knowledge of astronomy and cosmology.

## 4. A new explanation

There is little question that the kernel of Lemaître's 1927 paper was his brilliant hypothesis of a connection between the redshifts of the spiral nebulae and an expansion of space derived from the general theory of relativity. Indeed, it is generally accepted that it is this aspect of his paper that distinguishes his contribution from that of Alexander Friedman (Kragh and Smith 2003; Duerbeck and Seitter 2000; Nussbaumer and Bieri 2009 p113; Peebles 2015). However, while it is often assumed (see for example (van der Bergh 2011a; Way and Nussbaumer 2011)) that the astronomical data used by Lemaître to support his hypothesis were similar to the data published by Hubble in 1929, this assumption is not quite correct.

In his 1927 article, Lemaître derives an approximate relation between the fractional expansion of the cosmic radius $R$ and the velocities $v$ and distances $r$ of the spiral nebulae according to

$$\frac{R'}{R} = \frac{v}{cr} \qquad (1)$$

He then sets about obtaining an estimate for the rate of cosmic expansion using observational data for the redshifts and distances of the nebulae published by Gustav Strömberg and Edwin Hubble respectively (Strömberg 1925; Hubble 1926).[2] He notes first that the radial distances of the nebulae have been established using the method of apparent magnitude: *"The apparent magnitude m of these nebulæ can be found in the work of Hubble. It is possible to deduce their distance from it, because Hubble has shown that extragalactic nebulæ have approximately equal absolute magnitudes (magnitude = −15.2 at 10 parsecs, with individual variations ±2), the distance r expressed in parsecs is then given by the formula log r = 0.2m + 4.04"*.[3] Lemaître then calculates an approximate mean value for nebular distance, noting the large error associated with the method: *"One finds a mean distance of about $10^6$ parsecs, varying from a few tenths to 3,3 megaparsecs. The probable error resulting from the dispersion of absolute magnitudes is considerable. For a difference in absolute magnitude of ±2, the distance exceeds from 0.4 to 2.5 times the calculated distance. Moreover, the error is proportional to the distance"*.

Thus, taking redshifts and distances for 42 spiral nebulae from Strömberg and Hubble respectively, Lemaître extracts his estimate for cosmic expansion by dividing the mean velocity of the nebulae by the mean distance: *"Using the 42 nebulæ appearing in the lists of Hubble and Strömberg…and taking account of the proper velocity of the Sun…one finds a mean distance of 0,95 megaparsecs and a radial velocity of 600 Km/sec, i.e., 625 Km/sec at $10^6$ parsecs"*. Inserting the last two figures into the equation above he obtains

$$\frac{v}{cr} = \frac{625 x 10^5}{10^6 x 3.08 x 10^{18} x 3 x 10^{10}} = 0.68 x 10^{-27} cm^{-1} \qquad (2)$$

for the expansion rate of the cosmic radius.

Two aspects of this analysis are worth emphasizing. In the first instance, the nebular distances are taken from Hubble's publication of 1926 and thus almost all of them were estimated using the method of apparent magnitude. In this paper, Hubble himself is mindful of the many uncertainties associated with the method. Indeed, he notes early in the work that *"definite evidence as to distances and dimensions are restricted to six systems, including the Magellanic Clouds. The similar nature of the countless fainter nebulae has been inferred from*

---

[2] Almost all of the redshift data in Strömberg's paper is taken from the work of VM Slipher.
[3] All translations are taken from Jean-Pierre Luminet's translation of Lemaître's 1927 paper (Luminet 2013).

*the general principle of the uniformity of nature"* (Hubble 1926). Throughout the paper, Hubble describes the method of measuring nebular distance by apparent magnitude as a 'working hypothesis' and stresses that *"reliable values of distances, and hence of absolute magnitudes, are restricted to a very few of the brightest nebulae…..the number of known distance is too small to serve as a basis for estimates in the range in absolute magnitude among nebulae in general"* (Hubble 1926).[4] In the second instance, Lemaître does not claim that the data imply a linear relation between redshift and distance; instead he *predicts* the existence of such a relation from theory. This is seen most clearly in an important footnote to the section, where Lemaître notes that recent attempts by Knut Lundmark and Gustave Strömberg (Lundmark 1924; Strömberg 1925) to establish a relation between *v* and *r* indicate only a very weak correlation due to the uncertainties in nebular distance and suggests that a systematic error may be avoided by considering the ratio *v/r*:

> Some authors sought to highlight the relation between *v* and *r* and obtained only a very weak correlation between these two terms. The error in the determination of the individual distances is of the same order of magnitude as the interval covered by the observations and the proper velocity of nebulæ (in any direction) is large (300 Km/sec according to Strömberg), it thus seems that these negative results are neither for nor against the relativistic interpretation of the Doppler effect. The inaccuracy of the observations makes only possible to assume *v* proportional to *r* and to try to avoid a systematic error in the determination of the ratio *v/r*. Cf. Lundmark… and Strömberg, *l.c.*

In 1929, Edwin Hubble published the results of a detailed astronomical investigation of the relation between the redshifts of the nebulae and their distance, using the redshifts of Slipher and newly-derived estimates of nebular distances based on observations of individual stars within the nebulae (Hubble 1929). The results are reproduced in figure 1. As stated by Hubble: *"the results establish an approximately linear relation between the velocities and distances among nebulae for which velocities have been previously published"*. We note that the distances of the closest seven nebulae were estimated by observing Cepheid stars within the nebulae and employing Henrietta Leavitt's period-luminosity relation to estimate their distance; the next thirteen distances were estimated by observing the most luminous stars in nebulae and assuming an upper limit of absolute magnitude $M = -6.3$; the remaining four objects had distances assigned on the basis of the mean luminosities of the nebulae in a cluster.

---

[4] We note that most of the distances used in Hubble's paper are taken from the catalogue of Holeschek-Hopmann (Hubble 1926; Hopmann 1921).

Finally, the single cross represents a mean velocity/distance ratio for 22 nebulae whose distances could not be estimated individually (Hubble 1929). Even here, many commentators have noted that the quality and quantity of the data shown on Hubble's graph only marginally supported his conclusion of a linear relation between redshift and distance for the nebulae (Kragh 1996 p18; Longair 2006 p110; Nussbaumer and Bieri 2009 pp 115-116; Ostriker and Mitton 2013 p73; Peacock 2013; Grøn 2018). However, the graph marked an important turning point as the data were accepted by many theorists as the first evidence of an expansion of space (Smith 1979; Nussbaumer and Bieri 2006 pp 118-119; Kragh and Smith 2003). This conclusion was strengthened with the publication of a paper soon afterwards that extended the results to much larger distances and redshifts (Hubble and Humason 1931) and the result later became known as Hubble's law (Kragh and Smith 2003).

For comparison, a graph of the redshift/distance data used by Lemaître in 1927 is shown in figure 2; this graph was not shown in the original article but has been reconstructed by Hilmar Duerbeck and Waltraut Seitter from the data listed in Lemaître's paper (Duerbeck and Seitter 2000; Nussbaumer and Bieri 2009 pp 108-109). As Duerbeck and Seitter point out, a roughly linear trend of redshift versus distance can be discerned in this graph but the scatter in the data is much larger than for the case of Hubble's graph. More importantly, we note that the nebular distances of figure 2 were estimated using the method of apparent magnitude. Thus the data of figure 2 are quite different to Hubble's graph[5] and are more readily comparable with the earlier unsuccessful attempts mentioned above at establishing a redshift/distance relation for the nebulae by Lundmark and Strömberg, using redshifts from Slipher and nebular distances obtained using the method of apparent magnitude. The data of figure 2 can be viewed as a continuation of these investigations; indeed, we note that Duerbeck and Seitter calculate a correlation coefficient of 0.37, 0.23, 0.30 and 0.84 for the redshift/distance data of Lundmark, Strömberg, Lemaître and Hubble respectively (Duerbeck and Seitter 2000). In summary, it is clear that the data used by Lemaître in 1927 were not of the same calibre as the data of Hubble's graph of 1929. Thus we disagree with statements such as *"In 1929 Hubble repeats Lemaître's work with essentially the same data and obtains similar results"* (van der Bergh 2011a) or *"Two years later Hubble found the same velocity-distance relationship on observational grounds from practically the same observations that Lemaître had used"* (Way and Nussbaumer 2011). In our view, the observational data used by Lemaître in his paper of 1927 were of a preliminary nature because the nebular distances were established using a method that was prone to large errors; it is thus quite plausible that the data were not robust enough for Eddington (and others) to take Lemaître's new cosmological model seriously.

---

[5] Note that Hubble's graph is confined to more modest distances.

## 5. *Supporting evidence*

That the observational data used by Lemaître in his article of 1927 were of a preliminary nature can also be seen in his later actions and statements. For example, it has often been noted that in the 1931 translation, the entire data section of Lemaître's 1927 paper is replaced by a single sentence: "*From a discussion of available data, we adopt R'/R = 0.68x10$^{-27}$ cm$^{-1}$*" (Lemaître 1931a). It is now known that this editing was carried out by Lemaître himself (Livio 2013). Indeed, an excerpt from a letter from Lemaître to William Marshall Smart, editor of the *Monthly Notices*, clarifies the reason for the edit: "*I did not find it advisable to reprint the provisional discussion of radial velocities which is clearly of no actual interest*" (Lemaître 1931b). As more reliable estimates of the distances of the nebulae were available in the 1930s, why reproduce the 'provisional' data of his 1927 paper?

Many years later, Lemaître showed a similar attitude towards the data used in his article of 1927. In a letter to *Les Annales d'Astrophysique* in 1950, he commented: "*Naturellement, avant la décoverte et l'étude des amas de nébeleuses, il ne pouvait être question d'établir la loi de Hubble*" or "*Naturally, before the discovery and study of the clusters of nebulae, there was no question of establishing Hubble's law*" (Lemaître 1950). Similarly, in a review article of 1952, Lemaître wrote: "*Hubble and Humason established from observation the linear relation between velocity and distance which was expected for theoretical reasons and which is known as the Hubble velocity-distance relation*" (Lemaître 1952).

## 7. Conclusions

In conclusion, we are not convinced by conventional sociological explanations for Eddington's lack of interest in Lemaître's hypothesis of cosmic expansion when it was first proposed. We suggest instead a scientific explanation for the oversight. Extraordinary claims require extraordinary evidence and it is very plausible that the observational data cited by Lemaître in support of his model were not sufficiently robust to cause Eddington and others to consider expanding cosmologies. We note finally that by the time Eddington and his colleagues had embraced Lemaître's hypothesis of cosmic expansion, the brilliant young cleric had already embarked on his second great advance, the hypothesis of a universe with a hot, dense origin (Lemaître 1931c; Lambert 2015 pp 145-154; Kragh 2018).

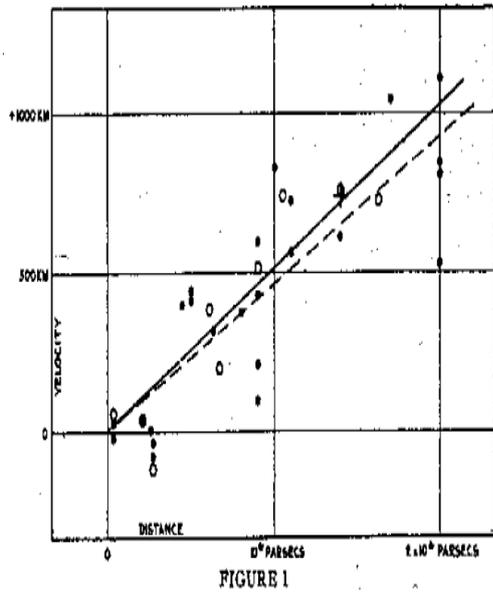

Figure 1. Hubble's graph of 1929, reproduced from (Hubble 1929).

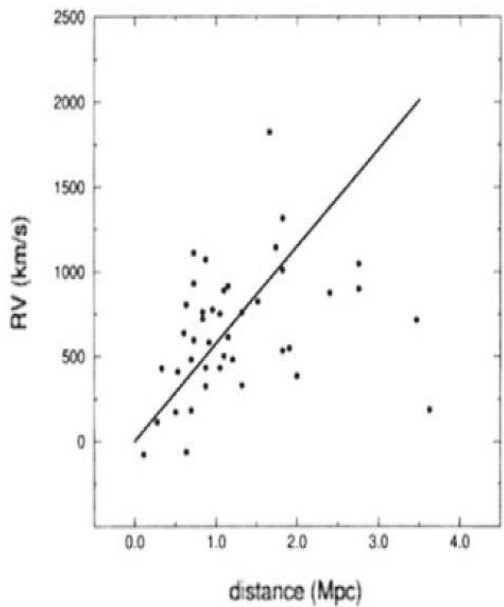

Figure 2. Reconstruction of Lemaître's data of 1927, reproduced from (Duerbeck and Seitter 2000).

**References**

Chandrasekhar, S. 1983. *Eddington: The Most Distinguished Astrophysicist of his Time.* Cambridge University Press, Cambridge.

de Sitter, W. 1917. On Einstein's theory of gravitation and its astronomical consequences. Third paper. *MNRAS* **78:** 3-28.

de Sitter, W. 1930. Proceedings of the RAS. *The Observatory* **53**: 37-39.

Duerbeck, H. W. and W.C. Seitter. 2000. In Hubble's shadow: Early research on the expansion of the universe. *Acta Hist. Astron*. **10**: 120-147.

Eddington, A.S. 1918. *Report on the Relativity Theory of Gravitation.* The Physical Society of London, Fleetway Press, London.

Eddington, A.S. 1920. *Space, Time and Gravitation: An Outline of the General Relativity Theory*. Cambridge University Press, Cambridge.

Eddington, A.S. 1923. *The Mathematical Theory of Relativity*. Cambridge University Press, Cambridge.

Eddington, A.S.1930. On the instability of Einstein's spherical world. *Month. Not. Roy. Astron. Soc.* **90**: 668-678.

Einstein, A. 1917. Kosmologische Betrachtungen zur allgemeinen Relativitätstheorie. *Sitz. König. Preuss. Akad*. 142-152. Or 'Cosmological considerations in the general theory of relativity' CPAE **6** (Doc. 43).

Friedman, A. 1922. Über die Krümmung des Raumes. *Zeit. Physik.* **10**: 377-386. Available in English translation as 'On the curvature of space' *General Relativity and Gravitation* **31**(12): 1991-2000 (1999).

Grøn, Ø. 2018. The discovery of the expansion of the universe. *Galaxies* **6**(4):132-142

Hopmann, J. 1921. Photometrische Untersuchungen von Nebelflecken. *Astr. Nach*. **214**: 28-

Hubble E. 1926. Extragalactic nebulae, *ApJ* **64**: 321-342.

Hubble, E. 1929. A relation between distance and radial velocity among extra-galactic nebulae. *Proc. Nat. Acad. Sci.* **15**: 168-173.

Hubble, E. and M.L. Humason 1931. The velocity-distance relation among extra-galactic nebulae. *ApJ* **74:** 43-93.

Kennefick, D. 2009. Testing relativity from the 1919 eclipse—a question of bias. *Physics Today* **62** (3): 37-40 https://doi.org/10.1063/1.3099578

Kragh, H. 1996. *Cosmology and Controversy*. Princeton University Press, Princeton.

Kragh, H. 2018. Georges, Lemaître, Pioneer of Modern Theoretical Cosmology. *Foundations of Physics* **48**(10): 1333-1348.

Kragh, H. and R. Smith. 2003. Who discovered the expanding universe? *Hist. Sci*. **41** :141-163.

Lambert, D. 2015. *The Atom of the Universe: The Life and Work of Georges Lemaître.* Copernicus Center Press, Krakow.

Lemaître, G. 1925. Note on de Sitter's universe. *J. Math. Phys*. **4:** 188-192.

Lemaître, G. 1927. Un univers homogène de masse constante et de rayon croissant, rendant compte de la vitesse radiale des nébuleuses extra-galactiques. *Ann. Soc. Sci. Brux.* **A47:** 49-59. Republished in English translation in *Gen. Rel. Grav*. **45**(8): 1619-1633 (Luminet 2013).

Lemaître, G. 1930. Letter to A.S. Eddington, February. Archives Lemaître, Université Catholique de Louvain.

Lemaître, G. 1931a. A homogeneous universe of constant mass and increasing radius, accounting for the radial velocity of the extra-galactic nebulae. *Mon. Not. Roy. Ast. Soc.* **91:** 483-490.

Lemaître, G. 1931b. Letter to W. Smart, March 9th. Correspondence of the Royal Astronomical Society.

Lemaître, G. 1931c. The beginning of the world from the point of view of quantum theory. *Nature* **127**: 706.

Lemaître, G. 1950. Compte rendu de P. Couderc; L'expansion de l'universe. *Ann d'Astro*. **13**: 344-345.

Lemaître, G. 1952. Clusters of nebulae in an expanding universe. *Mon. Not. Ast. Soc. S.A.* **11**: 110-117.

Livio, M. 2011. Lost in translation: mystery of the missing text solved. *Nature* **479**: 171-173.

Longair, M. 2006. *The Cosmic Century: A History of Astrophysics and Cosmology*. Cambridge University Press, Cambridge.

Lundmark, K. 1924. The determination of the curvature of space-time in de Sitter's world, *Mon. Not. Roy. Ast. Soc*. **84**: 747-757.

Luminet, J-P. 2013. Editorial note to 'A homogeneous universe of constant mass and increasing radius, accounting for the radial velocity of the extra-galactic nebulae'. *Gen. Rel. Grav*. **45**(8): 1619-1633.

McCrea, W. H. 1982. Recollections of Sir Arthur Eddington. *Contemp. Phys*. **23**: 531-540.

McVittie, G. 1967. Georges Lemaître. *Qt.J.Roy.Ast.Soc*. **8**: 294-297.

McVittie, G. 1978. Lemaître and Eddington. Interview with David DeVorkin, America Institute of Physics March 21st.

Merton, R. K. 1973. *The Sociology of Science: Theoretical and Empirical Investigations.* University of Chicago Press, Chicago.

Mitton, S. 2017. The Expanding Universe of Georges Lemaître. *Astronomy and Geophysics* **58**(2): 2.28-2.31.

Mitton, S. 2019. Private communication.

Nussbaumer, H. and L. Bieri. 2009. *Discovering the Expanding Universe*. Cambridge University Press, Cambridge.

Ostriker J.P. and S. Mitton. 2013. *Heart of Darkness: Unravelling the Mysteries of the Invisible Universe*. Princeton University Press, Princeton.

Peacock. J. A. 2013. Slipher, galaxies, and cosmological velocity fields. In *Origins of the Expanding Universe: 1912-1932* (Eds. M.Way and D. Hunter) Ast. Soc. Pacific. Conf. Series **471:** 3-25.

Peebles, P.J.E. 2015. Preface to *'The Atom of the Universe'* (Lambert 2015) pp 9-14.

Shaviv, G. 2011. Did Edwin Hubble plagiarize? Physics ArXiv:1107.0442

Slipher, V. M. 1917. Nebulae. *Proc. Am. Phil. Soc*. **56**: 403-409.

Smart, W.M., Temple, G., Spencer Jones H., Milne E.A, Dingle H. and H.N. Russell. 1945. Obituary: Sir Arthur Stanley Eddington. *The Observatory* **66**:1-12.

Smith, R.W. 1979. The origins of the velocity-distance relation. *J. Hist. Ast*. **10**: 133-164.

Strömberg, G. 1925. Analysis of Radial Velocities of Globular Clusters and Non-Galactic Nebulae, *ApJ* **61**: 35-3.

van den Bergh, S. 2011a. Discovery of the expansion of the universe. *J. Roy. Ast. Soc. Can.* **105**: 197.

van den Bergh, S. 2011b. The curious case of Lemaître's equation no. 24. *J. Roy. Ast. Soc. Can*. **105**: 151–152.

Vibert Douglas, A. 1958. *The Life of Arthur Stanley Eddington*. Nelson and Sons, London.

Way, M. and H. Nussbaumer. 2011. Lemaître's Hubble relationship. *Physics Today* **64**(8): 8.

Zuckerman, H. 1995. *Scientific Elite: Nobel Laureates in the United States*. Transaction Publishers, New York (reprint edition).